\documentclass[showpacs,preprint,pra]{revtex4}
\usepackage{amssymb}
\usepackage{amsmath}
\usepackage{amsfonts}
\usepackage{graphicx}
\usepackage{bm}
\begin{document}

\title{Geometric Phase of a Quantum Dot System in Nonunitary Evolution }

\author{Sun Yin\footnote{yinsun@sdu.edu.cn} and D. M. Tong\footnote{tdm@sdu.edu.cn}}
\affiliation{Department of physics, Shandong University, Jinan,
250100, P.R. China}

\date{\today}
\begin{abstract}
Practical implementations of quantum computing are always done in
the presence of decoherence. Geometric phase is useful in the
context of quantum computing as a tool to achieve fault tolerance.
Recent experimental progresses on coherent control of single
electron have suggested that electron in quantum dot systems is
promising candidate of qubit in future quantum information
processing devices. In this paper,  by considering a feasible
quantum dot model, we calculate the geometric phase of the quantum
dot system in nonuitary evolution and investigate the effect of
environment parameters on the phase value.
\end{abstract}
 \maketitle

The quantal geometric phase was first discovered by Berry
\cite{Berry} in 1984 in considering the quantum systems under cyclic
adiabatic evolution. It has aroused much attention of researchers
due to its importance. Since then the original notion of Berry phase
has been extended to a general concept of geometric phase for pure
states as well as for mixed states. The extension to pure states in
nonadiabatic cyclic evolution was developed by Aharonov and Anandan
\cite{Aharonov} in 1987, and that to pure states in nonadiabatic and
noncyclic evolution was done by Samuel and Bhandari \cite{Samuel} in
1988. Further generalizations and refinements, by relaxing the
constrains of adiabaticity, unitarity, and cyclicity of the
evolution, have since been carried out \cite{Mukunda,pati95}. While
all these extensions are of quantum systems in pure states,  Uhlmann
\cite{Uhlmann} was the first to address the geometric phases of
mixed states within the mathematical context of purification. A
physical definition of geometric phases for mixed states in unitary
evolution was put forward by Sj\"oqvist {\it et al.}
\cite{Sjoqvist2000} in 2000 based on quantum interferometry, and it
was recast in a kinematic description by Singh {\it et al.}
\cite{Singh2003}. The generalization of mixed geometric phases to
quantum systems in nonunitary evolution was given by Tong {\it et
al.} \cite{Tong2004} in 2004. More works on geometric phases related
to states for open systems may be seen in Refs. \cite{Carollo03}.

The geometric property of the geometric phase has stimulated many
applications. It has been found that the geometric phase plays
important roles in quantum phase transition, quantum information
processing, etc. \cite{Bohm}. The geometric phase shift can be fault
tolerant with respect to certain types of errors, thus several
proposals using NMR, laser trapped ions, etc. have been given to use
geometric phase to construct fault-tolerant quantum information
processer \cite{Falci2000}, and the fault-tolerant geometric quantum
computation gate has been demonstrated in experiments using NMR
\cite{Jones2000}.

Geometric phase is useful in the context of quantum computing as a
tool to achieve fault tolerance. Practical implementations of
quantum computing are always done in the presence of decoherence.
Fortunately, recent experimental progresses on coherent control of
single electron have suggested that electron in quantum dot systems
is a promising candidate of qubit in future quantum information
processing devices \cite{Loss1998}, because it has long spin
coherence time. This start us to investigate the geometric phase of
quantum dot systems in nonuitary evolution. In this paper, we
calculate the geometric phase of a feasible quantum dot model and
investigate the effects of the environment parameters to the phase
value.

\begin{figure}[htbp]
 \begin{center}
\includegraphics[width=4.5cm, height=2.5cm]{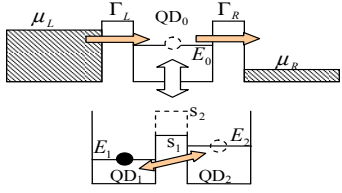}
 \end{center}
\caption{Illustration of the model. \label{model}}
\end{figure}
The model is illustrated as Fig.\ref{model}. Two quantum dots,
QD$_1$ and QD$_2$ , are coupled to each other with strength $s_1$.
An electron is trapped in the quantum dots and it tunnels between
the two quantum dots. Only one energy level is considered in each
quantum dot, and hence the electron and the two quantum dots
construct a two-level quantum system, a qubit. The environment of
the system consists of another quantum dot, QD$_0$, and two leads
connecting to QD$_0$. The left lead has higher chemical potential
than the right lead. Electrons can tunnel from the left lead to
QD$_0$ and then tunnel out to the right lead. For simplicity, we
assume that only one electronic state with the energy level $E_0$ in
QD$_0$ is correlated and $\mu_L>E_0>\mu_R$, where $\mu_L$ and
$\mu_R$ are the chemical potentials of the left lead and the right
lead respectively. Once there is an electron in QD$_0$, it will
affect the coupling between QD$_1$ and QD$_2$ by changing the
coupling strengthes from $s_1$ to $s_2$. This is an interesting
model, of which the relaxation and decoherence and quantum
measurement have been well studied \cite{Stace2004,Gurvitz2008}. The
model is easily performed in experiment, and it may play a potential
selection for geometric quantum computation of using quantum dot
systems.

Noting that the qubit system, comprising the trapped electron and
the two dots, is an open system being in mixed state, we use the
formula of geometric phases for mixed states in nonunitary evolution
given in Ref. \cite{Tong2004}. For an open quantum system, described
by the reduced density operator,
$\rho(t)=\sum_{k=1}^2\omega_k(t)|\phi_k(t)\rangle\langle\phi_k(t)|,~t\in[0,\tau]$
the geometric phase is given by the formula,
\begin{align}
\gamma(\tau)=\textrm{Arg}\bigg(\sum_{k=1}^2
\sqrt{\omega_k(0)\omega_k(\tau)}
\langle\phi_k(0)|\phi_k(\tau)\rangle e^{ -\int_0^\tau \langle
\phi_k(t) | \dot{\phi}_k(t) \rangle dt}\bigg), \label{GP}
\end{align}
where $\omega_k(t)$ is the $k-$th eigenvalue of the reduced density
matrix, $|\phi_k(t)\rangle$ is the corresponding eigenvector, and
$\tau$ is the total evolutional time.

In order to calculate the geometric phase of the qubit system, we
need to obtain the reduced density operator. The Hamiltonian of the
large system can be expressed as
\begin{align}
H=&H_s+H_e+H_i,\nonumber\\
H_s=&E_1a_1^{\dagger}a_1+E_2a_2^{\dagger}a_2+
s_1(a_1^{\dagger}a_2+a_2^{\dagger}a_1),\nonumber\\
H_e=&E_0c_0^{\dagger}c_0+\sum_lE_lc_l^{\dagger}c_l+\sum_r
E_rc_r^{\dagger}c_r +\sum_{l,r}(\Omega_lc_l^{\dagger}c_0
+\Omega_rc_0^{\dagger}c_r+\textrm{H.c.}),\nonumber\\
H_i=&(s_2-s_1) c_0^{\dagger}c_0(a_1^{\dagger}a_2+a_2^{\dagger}a_1).
\label{hamiltonian}
\end{align}
Here,  $H_s,~H_e,~H_i$ are the Hamiltonians corresponding to the
system itself, the environment and the interaction between the
system and its environment, respectively; $a_1^{\dagger}$ and
$a_2^{\dagger}$ $(a_1$ and $a_2)$ are the electron creation
(annihilation) operators in the two quantum dots; $c_l^{\dagger}$
and $c_r^{\dagger}$ ($c_l$ and $c_r$) are the electron creation
(annihilation) operators in the environment corresponding to the
left lead and the right lead respectively; $E_1$ and $E_2$ are the
energy level of QD$_1$ and QD$_2$;
$\Omega_l $ $(\Omega_r)$ is the coupling parameter of left (right)
lead with the quantum dot QD$_0$. For simplicity, we have considered
electrons as spinless fermions, and we have used $E_l$, $E_r$,
$\Omega_l $, and $\Omega_r$ to represent $E_{Ll}$, $E_{Rr}$,
$\Omega_{Ll} $, and $\Omega_{Rr}$ respectively.

The wave function of the large system, $|\Psi(t)\rangle$, satisfies
the Schr\"odinger equation,
$i\frac{d|\Psi(t)\rangle}{dt}=H(t)|\Psi(t)\rangle$. The reduced
density operator $\rho(t)$ may be expressed as the partial traces of
$|\Psi(t)\rangle\langle \Psi(t)|$ with respect to the environment
consisting of the quantum dot QD$_0$ and the two leads,
$\rho(t)=\text{tr}_{D_0}\varrho(t)$, where
$\varrho(t)=\text{tr}_{Ls} |\Psi(t)\rangle\langle \Psi(t)|$.
Following the method used in Refs. \cite{Gurvitz2008}, we may get
the equations of motion for the elements of density matrix
$\varrho(t)$. The bases of $\varrho(t)$ consists of four discrete
states, $|1\rangle\equiv|1,0,0\rangle$,
$|2\rangle\equiv|1,0,1\rangle$, $|3\rangle\equiv|0,1,0\rangle$,
$|4\rangle\equiv|0,1,1\rangle$, where $|n_1,n_2,n_3\rangle$ means
that there are $n_1,~n_2,~n_3$ electrons in QD$_1$, QD$_2$, QD$_0$
respectively. In the approximation of constant density of states,
let $\Gamma_L=2\pi |\Omega_{L} |^2\rho_{L}$ and $\Gamma_{R}=2\pi
|\Omega_{R}|^2\rho_{R}$, where $\rho_{L}$ ($\rho_{R}$)  is the
density of states for the left (right) lead, and $\Omega_L$
($\Omega_R$) denotes the constant coupling parameter $\Omega_l$
$(\Omega_r)$. $\Gamma_L$ ($\Gamma_R$) depicts the tunneling rate
between the left (right) lead and QD$_0$. In this case, the elements
$\varrho_{ij}$ of the density matrix $\varrho(t)$
satisfy\cite{note},
\begin{align}
\dot{\varrho}_{11}&=-\Gamma_L\varrho_{11}+\Gamma_R\varrho_{22}
-is_1(\varrho_{13}-\varrho_{31}),\nonumber\\
\dot{\varrho}_{22}&=-\Gamma_R\varrho_{22}+\Gamma_L\varrho_{11}
-is_2(\varrho_{24}-\varrho_{42}),\nonumber\\
\dot{\varrho}_{33}&=-\Gamma_L\varrho_{33}+\Gamma_R\varrho_{44}
-is_1(\varrho_{31}-\varrho_{13}),\nonumber\\
\dot{\varrho}_{44}&=-\Gamma_R\varrho_{44}+\Gamma_L\varrho_{33}
-is_2(\varrho_{42}-\varrho_{24}),\nonumber\\
\dot{\varrho}_{13}&=-i\epsilon_0\varrho_{13}-is_1(\varrho_{11}
-\varrho_{33})-\Gamma_L\varrho_{13}+\Gamma_R\varrho_{24},\nonumber\\
\dot{\varrho}_{24}&=-i\epsilon_0\varrho_{24}-is_2(\varrho_{22}
-\varrho_{44})-\Gamma_R\varrho_{24}+\Gamma_L\varrho_{13}.
\label{density}
\end{align}
The initial condition is taken as $\varrho_{ij}|_{t=0}=1$, for
$i=j=1$, or $0$, for all other $i,~j$, corresponding to the case
that the electron is in QD$_1$ and no electron is in QD$_0$. Here
$\epsilon_0=E_1-E_2$, is the energy difference of the energy levels
of QD$_1$ and QD$_2$. The elements $\rho_{ij}$ of the reduced
density matrix $\rho(t)$ of qubit can be then expressed as
\begin{align}
\rho_{11}=1-\rho_{22}=\varrho_{11}+\varrho_{22},~\rho_{12}=\rho_{21}^\ast=\varrho_{13}+\varrho_{24}.
\label{density-reduce}
\end{align}
Once the reduced density matrix is obtained, we can calculate its
eigenvalues $\omega_k(t)$ and eigenvectors $|\phi_k(t)\rangle$, and
we have
\begin{align}
&\omega_{1,2}(t)=\frac{1\pm\sqrt{(\rho_{11}
-\rho_{22})^2+4|\rho_{12}|^2}}{2},\nonumber\\
&|\phi_1(t)\rangle=\frac{1}{\sqrt{1+\frac{|\rho_{12}|^2}{(\omega_1
-\rho_{22})^2}}}\left[\begin{array}{c}1 \\
\frac{\rho_{21}}{\omega_1-\rho_{22}}\end{array}\right],\nonumber\\
&|\phi_2(t)\rangle=\frac{1}{\sqrt{1+\frac{|\rho_{12}|^2}{(\omega_2
-\rho_{11})^2}}}\left[\begin{array}{c} \frac{\rho_{12}}{\omega_2
-\rho_{11}}\\ 1
\end{array}\right].\label{omega1}
\end{align}
The initial condition taken above implies
$\omega_1(0)=1,~\omega_2(0)=0$, and $|\phi_1(0)\rangle=\Big[\begin{array}{c} 1\\
0\end{array}\Big], ~~|\phi_2(0)\rangle=\Big[ \begin{array} {c} 0\\
1\end{array}\Big].$

\begin{figure}[htbp]
 \begin{center}
\includegraphics[width=6.cm, height=4.5cm]{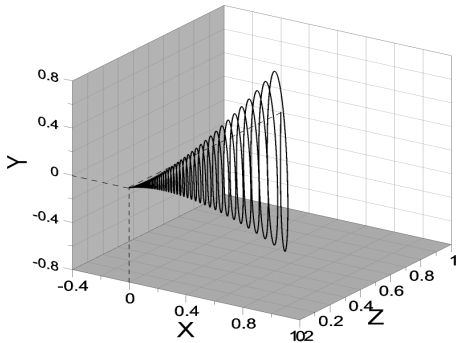}
 \end{center}
\caption{The Bloch sphere of the density matrix.
\label{BlochSphere}}
\end{figure}

The evolution of the system can be illustrated by the path traced in
Bloch sphere. The three-dimensional coordinates in the Bloch sphere
are $x=\rho_{12}+\rho_{21}$, $y=i(\rho_{12}-\rho_{21})$, and
$z=\rho_{11}-\rho_{22}$, respectively. By using the Four-order
Runge-Kutta method, we may numerically resolve the differential
equations in (\ref{density}) and obtain the value of the density
operator. Fig. \ref{BlochSphere} shows the path traced by the state
of the system, where the parameters are chosen as $\Gamma_L=1.0,~
\Gamma_R=2.0,~s_1=1.0,~s_2=0.5, ~\epsilon_0=-2.0$. Hereafter, we
take the parameter $s_1$ as the base unit.  All the other parameters
with energy dimension, such as  $\Gamma_L,~ \Gamma_R,~s_2$, are
measured by the unit $s_1$, and the time is measured by $1/s_1$. As
the time goes on, the path starts from $(0,0,1)$, which corresponds
to the state that the trapped electron is in QD$_1$, and moves
spirally to $(0,0,0)$, which corresponds to the state that the
electron has half probability in QD$_1$ and half in QD$_2$.

\begin{figure}[htbp]
 \begin{center}
\includegraphics[width=4.5cm, height=2.8cm]{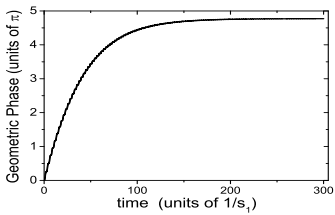}
 \end{center}
\caption{The geometric phase as a function of time. \label{GP-t}}
\end{figure}

Substituting Eq. (\ref{omega1}) into Eq. \eqref{GP}, we can
calculate the geometric phase of the system. It may be simply
expressed as
$\gamma(\tau)=i\int_0^\tau \langle \phi_1(t) |
\dot{\phi}_1(t)\rangle dt.$
To sketch out the changing tendency of the geometric phase, we
numerically calculate the geometric phase. The parameters are again
chosen as $\Gamma_L=1.0,~ \Gamma_R=2.0,~s_1=1.0,~s_2=0.5,
~\epsilon_0=-2.0$. The result is shown as Fig. \ref{GP-t}.
The geometric phase is usually put in region $[0,2\pi)$  (mod
$2\pi$). In order to show entirely the changing tendency of the
phase and express clearly the path dependence of the geometric
phase, here we give the schematic by using the calculated values
without making a $2\pi$-modulus. The recast of the results in
$[0,2\pi)$ is trivial.

From Fig. \ref{GP-t}, we find that the geometric phase is changing
as the time is going on, and it finally saturates to a constant
value. The saturation value is a characteristic value for a given
configuration of parameters, which may be simply called as the
characteristic geometric phase (CGP). This is consistent with the
`geometricity' of the geometric phase, that is, the geometric phase
is only dependent on the path traced by the state of the system, but
not on the dynamics. When evolutional time is small, the spiral path
has large spiral radius and the changing of the path is notable, and
thus the changing of the geometric phase is obvious too. With
evolutional time going on, the spiral radius of the path becomes
small and the changing rate of the spiral path are reduced, and
therefore the changing of the geometric phase will be reduced too.
The system will finally evolves to the point $(0,0,0)$, and from
then on the path will be little changing, and so does the geometric
phase.

In the model, there are three environment parameters  $s_2$,
$\Gamma_L$, $\Gamma_R$. We now investigate the effects of these
parameters on the phase values. For this, we will consider two kinds
of geometric phase values, the geometric phase corresponding to the
whole evolutional time, i.e., the CGP, and the geometric phase
corresponding to a special time interval $T$.

Firstly, we observe the effect of the parameters on the CGP.
\begin{figure}[htbp]
\begin{center}
\includegraphics[width=7.5cm, height=7.0cm]{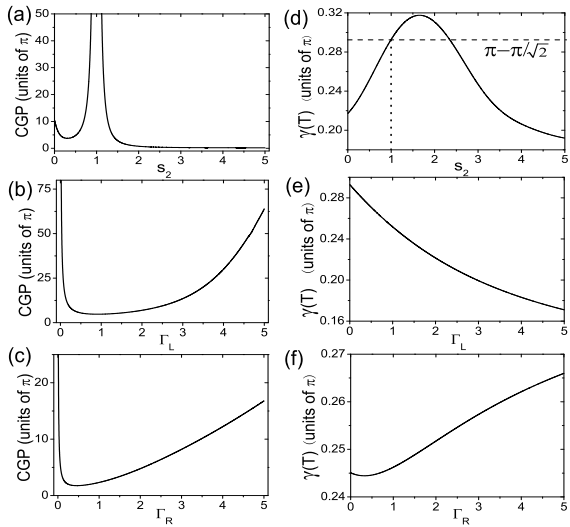}
 \end{center}
\caption{The geometric phases,  CGP and $\gamma(T)$, as functions of
the parameters $s_2$, $\Gamma_L$, and $\Gamma_R$. The parameters
except for the one taken as variable are chosen as $\Gamma_L=1.0,~
\Gamma_R=2.0,~s_1=1.0,~s_2=0.5, ~\epsilon_0=-2.0$.
\label{GP-parameter}}
\end{figure}
Fig. \ref{GP-parameter}(a) shows the effect of $s_2$ on CGP. From
the figure, we see that CGP is strongly dependent on the parameter
$s_2$. Specially, CGP is infinitely large at $s_2=s_1$. This is a
reasonable result, because $s_2=s_1$ means that the environment does
not affect the qubit system. In the case, the qubit is in the pure
state, which is evolving repeatedly along a closed circle in the
Bloch sphere, and CGP will accumulate infinitely as the time is
going on. However, as the parameter $s_2-s_1$ is becoming large from
zero, the value of the phase will reduce. The phase values will
approach to zero when $s_2-s_1$ is large enough. This may be
explained by the following argument. The larger $s_2-s_1$ means the
larger correlation between the environment and the qubit system,
which leads to the smaller spiral radius of the path traced by the
state of the system. When the environments' effect is stronger
enough, the path may approaches to a line directly from $(0,0,1)$ to
$(0,0,0)$ and the corresponding geometric phase will be near to
zero.

Figs. \ref{GP-parameter} (b) and (c) show the effect of parameters
$\Gamma_{L}$ and $\Gamma_{R}$ on CGP. From the figures, we find that
the two curves in the figures are similar. CGP becomes infinitely
large at $\Gamma_{L}=0$ or $\Gamma_{R}=0$, and it is also
approaching to infinity as $\Gamma_{L}$ or $\Gamma_{R}$ is going to
large values. These observations are consistent with the physical
construction in the model, as we have taken $\Gamma_{R}=2$ in Fig.
\ref{GP-parameter} (b) and $\Gamma_{L}=1$ in \ref{GP-parameter} (c).
Roughly speaking, when $\Gamma_L$ is small and $\Gamma_R$ is large,
electrons are hard to tunnel into QD$_0$ from the left lead but easy
to tunnel out of QD$_0$. There is nearly no electron staying in
QD$_0$ in all the time, i.e., the coupling between QD$_1$ and QD$_2$
is mainly $s_1$. The effect of the environment on the qubit is
negligible, and the qubit may be taken as a closed two-level system
with coupling strength $s_1$.  The picture of CGP corresponding to
the case is the left part of Fig. \ref{GP-parameter}(b) or the right
part of \ref{GP-parameter}(c). When $\Gamma_{L}$ is large and
$\Gamma_{R}$ is small, electrons are easy to tunnel into QD$_0$ from
the left lead but hard to tunnel out of QD$_0$. There is an electron
staying in QD$_0$ almost in all the time, i.e., the coupling between
QD$_1$ and QD$_2$ is dominated by $s_2$. The effect of the
environment on the qubit is only to change the coupling strength
between QD$_1$ and QD$_2$ from $s_1$ to $s_2$, and the qubit system
may be taken as a closed system but with coupling $s_2$. The picture
corresponding to this case is the right part of the curve in Fig.
\ref{GP-parameter}(b) or left part of the curve in Fig.
\ref{GP-parameter}(c). When $\Gamma_{L}$ and $\Gamma_{R}$ are in the
same order, the qubit is an open system in mixed state. The path
traced by the mixed state is a spiral curve and so corresponds to
finite values of CGP.

Secondly, we observe the effect of the parameters on the geometric
phase for the special time interval $T$. If there is no coupling
between the qubit and the environment, or $s_2=s_1$, the qubit
system will be in a pure state and it will evolve from the initial
state $(0,0,1)$ back to itself after a time interval $T$, making up
a closed circle in the Bloch sphere. In the case where
$\epsilon_0=-2$ and $s_1=1$, we have $T=\pi/\sqrt{2}$, and the
geometric phase corresponding to the closed cycle is
$\gamma(T)=\pi-\pi/\sqrt{2}$. However, if $s_2\ne s_1$, the path
traced by the state in the Bloch sphere will become an unclosed
curve and the geometric phase $\gamma(T)$ will be changed under the
effect of the environment. Therefore, $\gamma(T)$ may be used to
describe the effect of the environment on geometric phase in an
finite time, during which the pure state evolves one circle. Figs.
\ref{GP-parameter}(d),  \ref{GP-parameter}(e) and
\ref{GP-parameter}(f) show the effect of parameters $s_2$,
$\Gamma_L$ and $\Gamma_R$ on $\gamma(T)$, respectively. The curves
in the figures may be explained by applying a similar discussion as
above.

In conclusion, we have calculated the geometric phase of a feasible
quantum dot model and investigate the effects of the environment
parameters to the phase value. Here, we not only presented the
parameters's effect on the characteristic geometric phase, which
corresponding to the whole evolutional time, but also studied their
effect on the geometric phase in a finite time interval $T$, defined
by using pure state without the effect of environment. The approach
of calculating the geometric phase in the paper is reliable. While
the other approaches of defining the geometric phase of open systems
have met criticisms \cite{Ericsson2003}, the kinematic approach used
in the paper has been widely applied to investigate the open systems
in various environments \cite{Hamonic}. Our investigation on
geometric phase is helpful to completely understand the properties
of the quantum dot system.

This work is supported by NSF of China under Grant Nos.10675076,
10875072 and 10804062.


\begin{references}
\bibitem{Berry} M.V. Berry,
Proc. R. Soc. London Ser. A {\bf 392}, 45 (1984).
\bibitem{Aharonov} Y. Aharonov and J. Anandan,
Phys. Rev. Lett. {\bf 58}, 1593 (1987); J. Anandan and Y. Aharonov,
Phys. Rev. D {\bf 38}, 1863 (1988).
\bibitem{Samuel} J. Samuel and R. Bhandari,
Phys. Rev. Lett. {\bf 60}, 2339 (1988).
\bibitem{Mukunda} N. Mukunda and R. Simon,
Ann. Phys. (N.Y.) {\bf 228}, 205 (1993).
\bibitem{pati95} A.K. Pati,
Phys. Rev. A {\bf 52}, 2576 (1995); J. Phys. A {\bf 28}, 2087
(1995).
\bibitem{Uhlmann} A. Uhlmann,
Rep. Math. Phys. {\bf 24}, 229 (1986); Lett. Math. Phys.{\bf 21},
229 (1991).
\bibitem{Sjoqvist2000} E. Sj\"oqvist {\it et al.}, Phys. Rev. Lett. {\bf
85}, 2845 (2000).
\bibitem{Singh2003}K. Singh {\it et al.}, Phys. Rev. A {\bf 67}, 032106 (2003).
\bibitem{Tong2004} D.M. Tong {\it et al.},
Phys. Rev. Lett. {\bf 93}, 080405 (2004).
\bibitem{Carollo03} M. Ericsson {\it et al.}, Phys. Rev. A {\bf 67}, 020101(R)
(2003); A. Carollo {\it et al.}, Phys. Rev. Lett. {\bf 90}, 160402
(2003);  K. P. Marzlin, S. Ghose, and B.C. Sanders, {\it ibid.} {\bf
93}, 260402 (2004); X. X. Yi, L. C. Wang, and T. Y. Zheng,
 {\it ibid.} {\bf 92}, 150406 (2004).
\bibitem{Bohm} A. Bohm {\it et al.}, {\it The Geometric Phase in Quantum Systems} (Springer,
New York, 2003); S.-L. Zhu, Phys. Rev. Lett. {\bf 96}, 077206
(2006); A.C.M. Carollo and J.K. Pachos, {\it ibid.} {\bf 95}, 157203
(2005).
\bibitem{Falci2000} G. Falci {\it et al.}, Nature {\bf 407}, 355 (2000); X.-B. Wang and M. Kerji, Phys. Rev. Lett. {\bf
87}, 097901 (2001); L.-M. Duan, J.I. Cirac and P. Zoller, Science
{\bf 292}, 1695 (2001).
\bibitem{Jones2000} J.A. Jones {\it et al.},
Nature {\bf 403}, 869 (2000).
\bibitem{Loss1998} D. Loss and D.P. DiVincenzo, Phys. Rev. A {\bf
57}, 120 (1998); J.R. Petta {\it et al.}, Science {\bf 309}, 2180
(2005); D. Press {\it et al.}, Nature  {\bf 456}, 218 (2008); P.
San-Jose {\it et al.}, Phys. Rev. B {\bf 77}, 045305 (2008).
\bibitem{Stace2004}T.M. Stace {\it et al.},
Phys. Rev. B {\bf 70},205342 (2004).
\bibitem{Gurvitz2008} S.A. Gurvitz and D. Mozyrsky, Phys. Rev. B {\bf 77},
075325 (2008); S.A. Gurvitz and G.P. Berman,  {\it ibid.} {\bf 72},
073303 (2005); T. Gilad and S.A. Gurvitz, Phys. Rev. Lett. {\bf 97},
116806 (2006).
\bibitem{note}  $|\Psi(t)\rangle$ may be expanded in the picture of the
creation and annihilation operators $( a_1^{\dagger},~ a_1,
~a_2^{\dagger},~a_2, ~c_l^{\dagger},~c_l,~c_r^{\dagger},~c_r)$
applying on the ``vacuum'' state of the large system with all the
levels in the two leads being filled with electrons up to the Fermi
levels. Substituting the expansion of $|\Psi(t)\rangle$ into the
Schr\"odinger equation, one may get the differential equations
satisfied by the expansion amplitudes, from which Eq.\eqref{density}
can be derived by tracing out the freedoms of the two leads. Please
refer to \cite{Gurvitz2008} for details.
\bibitem{Ericsson2003} A. Bassi and E. Ippoliti, Phys. Rev. A {\bf
73}, 062104 (2006); M. Ericsson {\it et al.}, Phys. Rev. Lett. {\bf
91}, 090405 (2003).
\bibitem{Hamonic} X.X. Yi, L.C. Wang and W. Wang, Phys. Rev. A
{\bf 71}, 044101 (2005); X.X. Yi {\it et al.}, {\it ibid.} {\bf 73},
052103 (2006); A.T. Rezakhani and P. Zanardi, {\it ibid.} {\bf 73},
052117 (2006); F.C. Lombardo and P.I. Villar, {\it ibid.} {\bf 74},
042311 (2006); X.X. Yi and W. Wang, {\it ibid.} {\bf 75}, 032103
(2007); J. Dajka, M. Mierzejewski and J. Luczka, J. Phys. A {\bf
41}, 012001 (2008); J. Dajka and J. Luczka, {\it ibid.} {\bf 41},
442001 (2008); S. Banerjee and R. Srikanth, Euro. Phys. J. D {\bf
46}, 335 (2008).
\end{references}
\end{document}